\documentclass[aps,prb,twocolumn,10pt,longbibliography]{revtex4-1}
\usepackage{amsmath}
\usepackage{amsfonts}
\usepackage{amssymb}{\tiny }
\usepackage{mathtools}
\usepackage{times}
\usepackage{hyperref}
\usepackage{bm}
\usepackage{cases}
\usepackage[version=4]{mhchem}
\usepackage{graphicx}
\usepackage[caption=false]{subfig}
\usepackage{xcolor}
\allowdisplaybreaks%

\begin{document}
\title{Thermal Hall signatures of non-Kitaev spin liquids in honeycomb Kitaev materials}
\author{Yong Hao Gao$^{1}$}
\author{Ciar\'{a}n Hickey$^{2}$}
\author{Tao Xiang$^{3,4}$}
\author{Simon Trebst$^{2}$}
\author{Gang Chen$^{5}$}
\affiliation{$^{1}$State Key Laboratory of Surface Physics and Department of Physics, 
Fudan University, Shanghai 200433, China}
\affiliation{$^{2}$Institute for Theoretical Physics, University of Cologne, 50937 Cologne, Germany}
\affiliation{$^{3}$Institute of Physics, Chinese Academy of Sciences, Beijing 100190, China}
\affiliation{$^{4}$University of Chinese Academy of Sciences, Beijing 100049, China}
\affiliation{$^{5}$Department of Physics and Center of Theoretical and Computational Physics, 
The University of Hong Kong, Pokfulam Road, Hong Kong, China}

\date{\today}

\begin{abstract}
Motivated by the recent surge of field-driven phenomena discussed for Kitaev materials,
in particular the experimental observation of a finite thermal Hall effect and
theoretical proposals for the emergence of additional spin liquid phases beyond the 
conventional Kitaev spin liquid, we develop a theoretical understanding of the 
thermal Hall effect in honeycomb Kitaev materials in magnetic fields. Our focus 
is on gapless U(1) spin liquids with a spinon Fermi surface, which have been 
shown to arise as field-induced phases. We demonstrate that in the presence 
of symmetry-allowed, second-neighbor Dzyaloshinskii-Moriya interactions these 
spin liquids give rise to a finite, non-quantized, thermal Hall conductance 
in a magnetic field. The microscopic origin of this thermal Hall effect can 
be traced back to an interplay of Dzyaloshinskii-Moriya interaction and 
Zeeman coupling, which generates an internal U(1) gauge flux that twists 
the motion of the emergent spinons. We argue that such a non-quantized 
thermal Hall effect is a generic response in Kitaev models for a range 
of couplings.
\end{abstract}
\maketitle

\section{Introduction}
\label{sec1}

The first experimental observation of a quantum Hall effect 
in two-dimensional electron systems~\cite{Klitzing1980} 
has proved to be a scientific revolution, with its exact 
quantization of Hall resistance raising measurement standards to 
unprecedented levels of precision~\cite{Klitzing2019}.
It has also served as a blueprint for the interplay 
between experimental breakthroughs and deep conceptual progress
on the theory side. For the integer quantum Hall effect, 
it has been the seminal introduction of topological invariants~\cite{Thouless1982} 
to explain the quantization of conductance. 
For the subsequent fractional quantum Hall effect~\cite{Tsui1982}, 
it has been the theoretical concepts of emergence and fractionalization~\cite{Laughlin1983}. 
The observation of the quantum spin Hall effect~\cite{Koenig2007} 
has marked the birth of the topological insulator~\cite{Moore2010}. 
It is therefore that the more recent observation of a 
half-integer quantized thermal Hall effect~\cite{Kasahara2018,Banerjee2018} 
has caught the imagination of experimentalists and theorists alike.

In one of these experiments~\cite{Kasahara2018}, a thermal Hall effect 
is observed in crystalline samples of RuCl$_3$ -- a Mott insulator, 
in which the electronic degrees of freedom are frozen out~\footnote{
The charge gap for RuCl$_3$ has been reported to be within $0.22$ to $1.2$~eV in experimental studies 
\cite{PhysRevB.94.161106,PhysRevB.90.041112,PhysRevB.93.075144,PhysRevLett.117.126403}.
}
and the heat transport~\cite{PhysRevLett.118.187203} must be facilitated through charge-neutral modes. 
With the thermal conductance being quantized at a half-integer value, 
this points to the striking possibility of a Majorana
fermion edge current forming in these systems. 
On the theoretical side, this is rationalized by the designation
of RuCl$_3$ as a ``Kitaev material''~\cite{Trebst2017}--special types 
of spin-orbit assisted Mott insulators~\cite{WitczakKrempa2014,Rau2016}, 
in which local spin-orbit entangled ${j=1/2}$ 
moments~\cite{Khaliullin2005,PhysRevB.78.094403,PhysRevLett.101.076402} 
form that are subject to bond-directional exchanges~\cite{PhysRevLett.102.017205} 
familiar from the celebrated Kitaev model~\cite{KITAEV20062}.
The appeal of making such a direct connection to this elementary 
spin model comes from the fact that the latter exhibits a
number of quantum spin liquid ground states \cite{Balents2010,Savary2016}. 
Out of these, the field-induced, gapped topological 
spin liquid, often simply referred to as ``Kitaev spin liquid'', 
is a chiral spin liquid with gapless Majorana edge modes. As such
it appears to be a natural fit to explain the quantized thermal 
Hall effect in RuCl$_3$, in particular after considering the subtle
interplay of gapless Majorana and phonon modes in such a chiral 
spin liquid~\cite{PhysRevX.8.031032,PhysRevLett.121.147201}.

\begin{figure}[b]
	\centering
	\includegraphics[width=7cm]{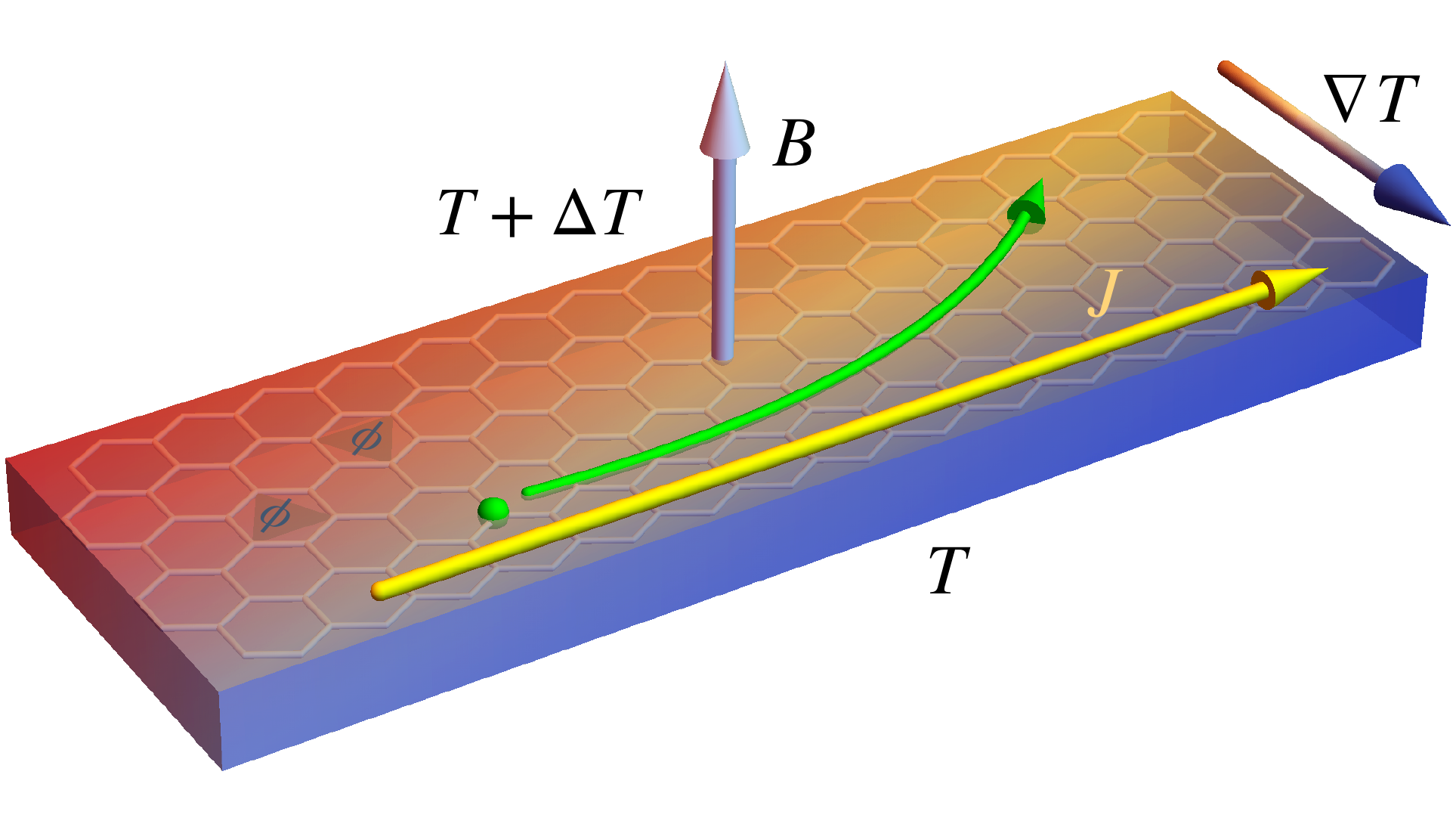}
	\caption{Schematic illustration of the {\bf thermal Hall effect of charge-neutral spinons} 
	arising for a field-induced U(1) spin liquid with a spinon Fermi surface 
	in honeycomb Kitaev materials in an external magnetic field.}
	\label{fig1}
\end{figure}

The observation of a finite, but non-quantized thermal Hall effect 
is an even more general, though still unusual phenomenon, which has 
been reported not only for a broad temperature and magnetic field 
range for RuCl$_3$~\cite{PhysRevLett.120.217205,PhysRevB.99.085136} 
(in addition to the quantized regime), but also a number of other 
spin liquid candidate materials such as the kagome magnets volborthite~\cite{Watanabe8653} 
Cu$_3$V$_2$O$_7$(OH)$_2 \cdot 2$H$_2$O and kapellasite~\cite{PhysRevLett.121.097203}
CaCu$_3$(OH)$_6$Cl$_2 \cdot 0.6$H$_2$O, as well as the pyrochlore 
spin ice material~\cite{PhysRevLett.115.106603} Tb$_2$Ti$_2$O$_7$.
This points to an alternative microscopic origin of charge-neutral 
thermal transport beyond the one sketched above for the gapped, 
chiral spin liquid, which always leads to a quantized Hall 
effect~\cite{PhysRevX.8.031032,PhysRevLett.121.147201}. 
Indeed, as some of us have recently pointed out in the context of 
kagome spin liquids~\cite{Chen1901.01522}, there is the possibility 
that even a {\em gapless} quantum spin liquid can exhibit a finite 
thermal Hall conductivity. The microscopic mechanism at play involves 
an interplay between the emergent, charge-neutral spinon degrees of 
freedom and certain types of Dzyaloshinskii-Moriya interactions that 
lead to a twist in the spinon motion, thereby allowing for a transverse
heat flow.

It is the purpose of the manuscript at hand to generalize this idea of 
emergent spinons mediating a charge-neutral, transverse thermal Hall 
current to honeycomb Kitaev materials. The reason to do so is motivated 
not only by the experimental observations~\cite{PhysRevLett.120.217205,PhysRevB.99.085136} 
for RuCl$_3$ discussed above, but also recent reports of the emergence 
of a field-driven gapless U(1) spin liquid~\cite{Hickey2019,Lu1809.08247,He1809.09091} 
in antiferromagnetic~\cite{PhysRevB.97.241110} Kitaev magnets
paired with {\em ab initio} modelling for the honeycomb Kitaev 
materials RuCl$_3$, Na$_2$IrO$_3$, and $\alpha$-Li$_2$IrO$_3$, 
which report, in unison, a variety of additional interaction terms beyond a dominant
bond-directional Kitaev exchange~\cite{Rau2016,PhysRevB.93.214431,Winter_2017}.    
We demonstrate that in the presence of a general, symmetry allowed, 
next-nearest neighbor Dzyaloshinskii-Moriya (DM) interaction
the emergent, gapless spinon degrees of freedom of such a field-driven 
U(1) spin liquid will indeed contribute to a transverse heat flow.
This is a priori not obvious, since the net internal flux in every 
unit cell vanishes and there are thus no obvious ``spinon Landau levels'' 
or quantum oscillations. However, we argue along the lines of 
Ref.~\onlinecite{Chen1901.01522} that the spinons hopping between 
the second neighbor sites will still experience an induced internal 
gauge flux. We explicitly demonstrate that this mechanism generates 
a non-trivial Berry curvature and thereby allows for a significant 
spinon thermal Hall effect, as schematically illustrated in Fig.~\ref{fig1}.
In reverse, this leads us to conclude that the observation of a finite, 
non-quantized thermal Hall conductance in honeycomb Kitaev materials 
would point towards the possibility of ``non-Kitaev spin liquid'' 
regimes in these materials. We further substantiate this reasoning 
by considering a gapless Dirac spin liquid, a third possible spin liquid 
scenario besides the chiral spin liquid and the gapless spinon Fermi 
surface U(1) spin liquid, for which we arrive at a similar conclusion.

The discussion in the remainder of this paper is structured as follows. 
We begin in Sec.~\ref{Sec:sec2} with a detailed review of the symmetry 
allowed microscopic spin model for the honeycomb iridates. 
In Sec.~\ref{Sec:sec3}, we consider the U(1) spin liquid state 
with a neutral spinon Fermi surface and numerically estimate the spinon thermal 
Hall conductivity within linear response theory. In Sec.~\ref{Sec:sec4}, 
we inspect an alternate scenario of a Dirac spin liquid. We conclude in Sec.~\ref{Sec:sec5} 
with a discussion of the results and the relevance of thermal transport measurements in the 
honeycomb Kitaev materials, including $\alpha$-RuCl$_3$
and other Kitaev materials such as H$_3$LiIr$_2$O$_6$. 
Technical details of our calculations and some further 
aspects of the materials are presented in the Appendices.

\section{Spin-orbit coupling and the Spin Model}
\label{Sec:sec2}

To set the stage for our discussion, we start by briefly providing a comprehensive and self-contained review of the microscopic physics of honeycomb Kitaev iridates. We put a particular focus on the symmetry-allowed exchange interactions beyond a bare, bond-directional Kitaev coupling and discuss the explicit form of the spin-orbit induced DM interaction.
More extensive introductions to this basic microscopic physics may be found in early theory works on the 
iridates~\cite{Khaliullin2005,PhysRevLett.102.017205,PhysRevLett.105.027204}, 
as well as more recent review articles~\cite{Trebst2017,doi:10.1146/annurev-conmatphys-033117-053934}. 

\subsection{Spin-orbit coupling and the derivation of the superexchange interaction}
\label{sec:sec2a}

\begin{figure}[b]
	\centering
	\includegraphics[width=4.1cm]{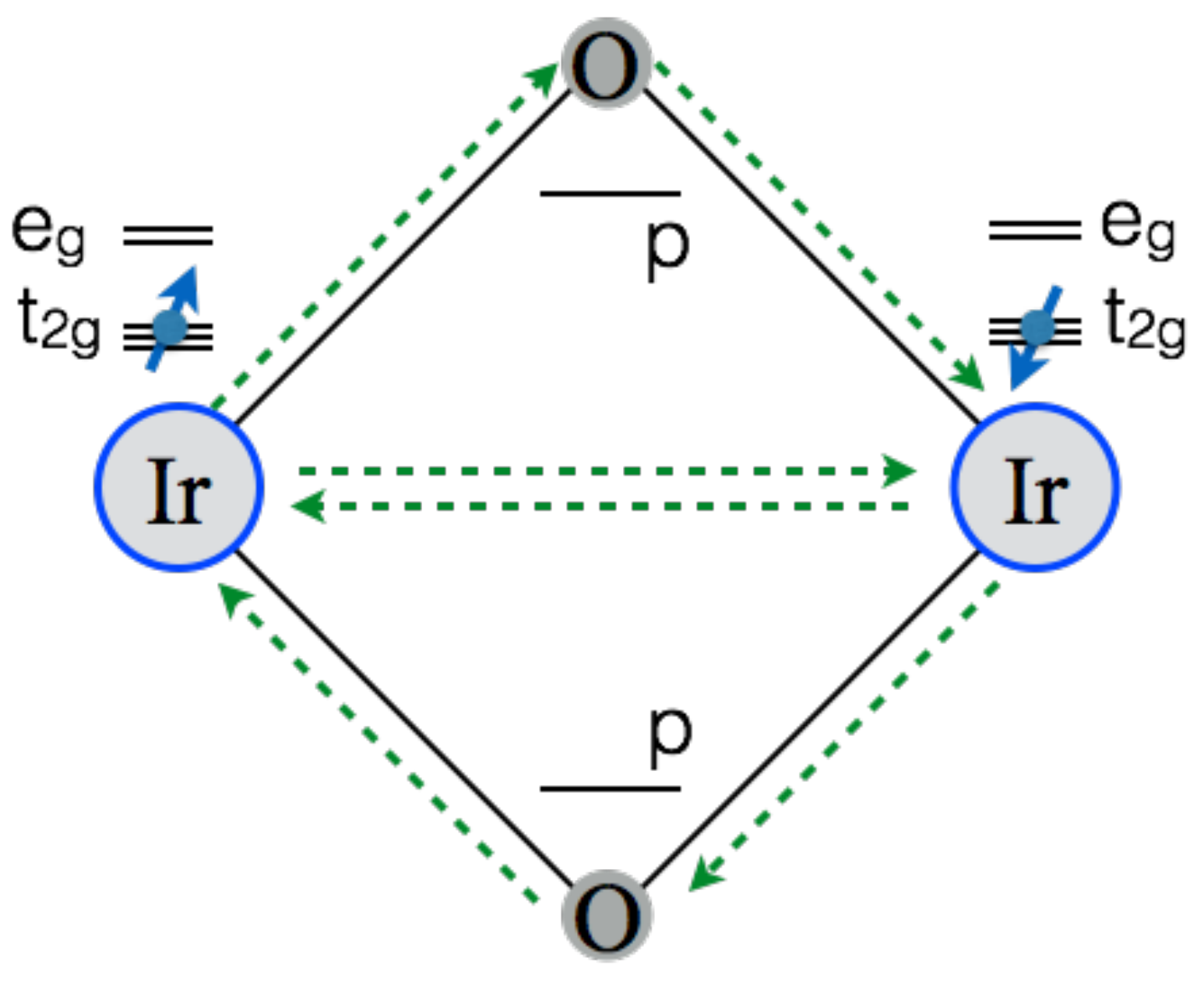} \hspace{0.2cm}
	\includegraphics[width=4.1cm]{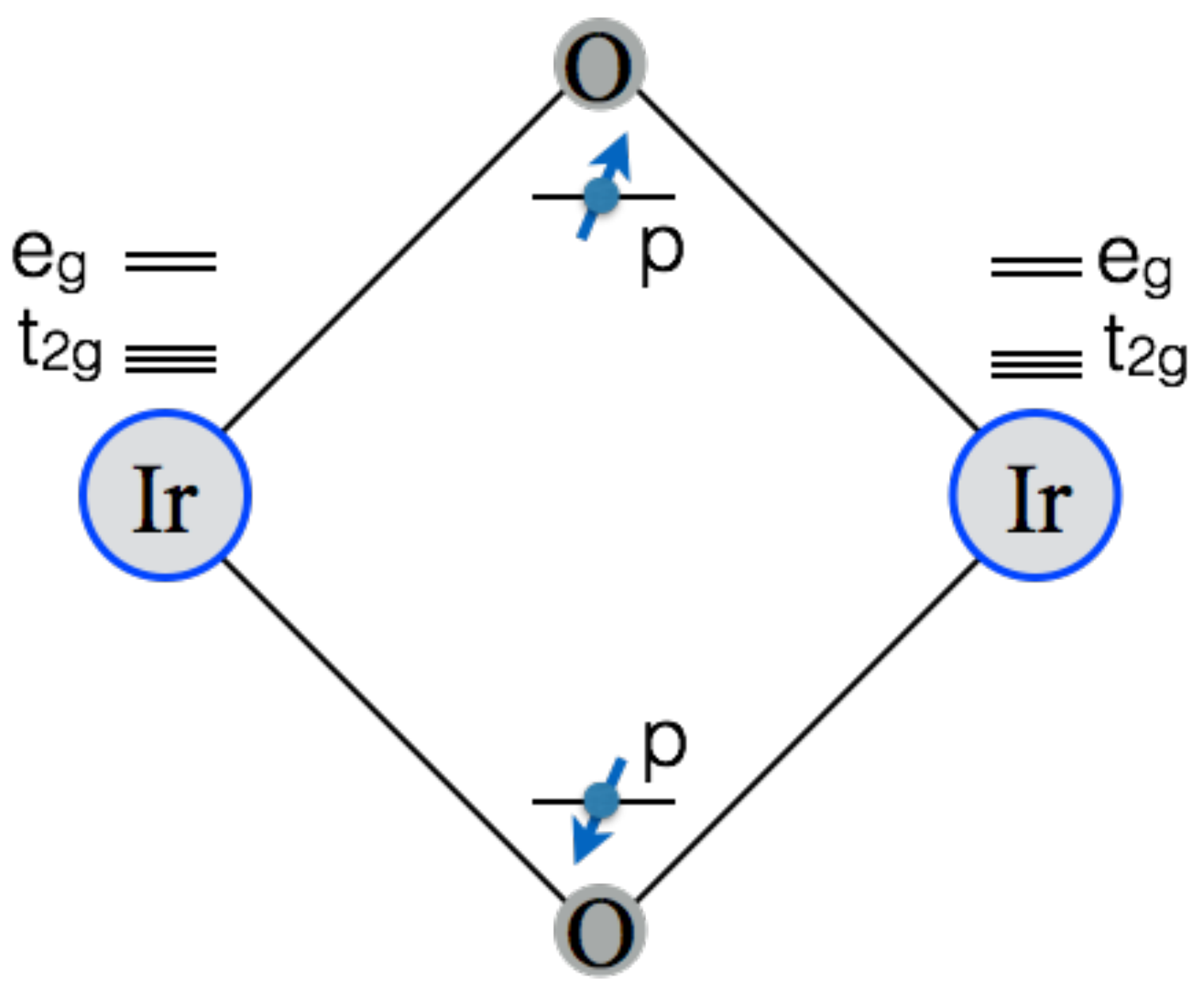}
	\caption{{\bf Microscopics of honeycomb Kitaev materials.}
	(a) The superexchange paths for different approximation schemes. 
	(b) The intermediate electron/hole configuration for the superexchange through 
	the oxygen atoms. }
	\label{exchange}
\end{figure}

The iridium atom has an atomic number ${Z=77}$ and thus the $5d$ electrons   
of the iridium experience a much stronger spin-orbit coupling than the $3d$ 
electrons of transition metal ions. The full Hamiltonian of the iridates is 
given as~\cite{PhysRevB.78.094403}
\begin{eqnarray}
H  & = &  H_{\text{Ir-O}} + H_{\text{Ir-Ir}} + H_{\text{soc}} + H_{\text{split}} 
\nonumber \\
&&   \quad\quad\quad  + H_{\text{Ir-corr}} + H_{\text{O-corr}},
\label{eq1}
\end{eqnarray}
where $H_{\text{Ir-O}}$ describes the hopping between the Ir $5d$ ($t_{2g}$) orbitals 
and the O $p$ orbitals, $H_{\text{Ir-Ir}}$ is the direct hopping between the Ir $t_{2g}$ 
electrons, $H_{\text{soc}}$ is the atomic spin-orbit coupling (SOC)
of the Ir $t_{2g}$ electrons, $H_{\text{split}}$ describes the crystal field splitting 
within $t_{2g}$ orbitals, $H_{\text{Ir-corr}}$ is the electron correlation for the 
Ir $t_{2g}$ electrons and is parametrized by various Kanamori interactions, and 
$H_{\text{O-corr}}$ is the electron correlation within the O $p$ orbitals. In this 
extended Hubbard model, we have already made the approximation to truncate the 
upper $e_g$ orbitals. The $e_g$-$t_{2g}$ splitting is of the order of a couple eVs, 
and ignoring $e_g$ orbitals is a good approximation for the $5d$ materials.
This extended Hubbard model describes almost all of the iridates of interest. 
The Ir-O-Ir complex that often occurs in various iridates is given in Fig.~\ref{exchange}.

This extended Hubbard model is similar to the three-band model for cuprates
except that here multiple $t_{2g}$ orbitals are involved and a strong spin-orbit 
coupling is present. In the context of cuprates, where the concept of the charge 
transfer insulator is relevant, the further reduction of the three-band model 
to the single-band (``$t$-$J$'' type) model is made possible through the 
observation of the Zhang-Rice singlets and the virtual sharing of the 
Zhang-Rice singlets between neighboring Cu-O complexes.

Since we are interested in the Mott insulating regime, all the electrons 
are localized on the ionic sites and form a local moment. 
In the single ion limit, there are five $5d$ electrons on the 
$t_{2g}$ shell, and SOC is active at the linear order. 
In the single-electron basis, the atomic SOC appears as 
\begin{eqnarray}
H_{\text{soc}} &=& \sum_i \lambda\, {\boldsymbol L}_i \cdot {\boldsymbol S}_i
\\
&\Rightarrow & - \sum_i \lambda\, {\boldsymbol l}_i \cdot {\boldsymbol S}_i,
\end{eqnarray}
where ${\boldsymbol L}_i$ operates on the five $5d$ orbitals (including both $e_g$ and
$t_{2g}$) with ${L=2}$, and ${\boldsymbol l}$ operates within the lower $t_{2g}$ triplets,
and more importantly, ${l=1}$. Thus, the SOC favors a lower quadruplet with ${j=3/2}$
and a upper doublet with ${j=1/2}$. Four electrons would completely fill the lower 
quadruplets, and the remaining electron occupies the upper ${j=1/2}$ (Kramers) doublet
and gives rise to the effective spin-1/2 local moment. In the absence of
the further splitting among the $t_{2g}$ shell, the wavefunctions of the ${j=1/2}$ 
states are given as 
\begin{eqnarray}
|j^z=\uparrow \rangle  &=& 
\frac{1}{\sqrt{3}} [ -i | xz, \downarrow \rangle + | yz, \downarrow \rangle  
                                                 + | xy, \uparrow \rangle  ],
                                                 \\
|j^z=\downarrow \rangle  &=&    
\frac{1}{\sqrt{3}} [+ i | xz, \uparrow \rangle + | yz, \uparrow \rangle  
                                                 - | xy, \downarrow \rangle  ].   
\end{eqnarray}
Remarkably, because of the involvement of the orbitals, the Land\'{e} factor 
is ${g=-2}$ instead of the usual ${g=2}$ for the spin-only local moments. 
However, this sign difference cannot be measured in the magnetic susceptibility 
in which $g$ enters at an even power. In any real material, there exist small splittings 
among the $t_{2g}$ orbitals. Such splittings lead to further modifications of the effective 
${j=1/2}$ wavefunction, and change the $g$ factor continuously from $-2$ 
to $+2$ for the limit that orbital degeneracy is completely broken and SOC 
is quenched. So the combination of  SOC and $t_{2g}$ splitting 
could generate a rather small magnetic moment. This can probably account 
for the small magnetic moments~\cite{PhysRevB.88.220414} 
of Na$_2$IrO$_3$ and Li$_2$IrO$_3$.

As the microscopic Hamiltonian in Eq.~\eqref{eq1} 
contains many different interactions and couplings,
there exist various approximation schemes to deal with 
this extended Hubbard model. One approximation is to 
keep the direct Ir-Ir hopping, the on-site SOC and 
the Hubbard-$U$ correlation, {\sl i.e.}
\begin{eqnarray}
H_{\text{A}_1} = H_{\text{Ir-Ir}} + H_{\text{soc}} + H_{\text{Ir-corr}}.
\end{eqnarray}
One then projects the model onto the ${j=1/2}$ manifold. 
Remarkably, one obtains an apparent isotropy even though 
the effective spin itself contains a substantial orbital 
component, and the resulting spin model is a Heisenberg 
model~\cite{PhysRevB.78.094403,PhysRevLett.102.017205}.
Another approximation is to consider the superexchange 
through the oxygen with the starting Hamiltonian as 
\begin{eqnarray}
H_{\text{A}_2} = H_{\text{Ir-O}}  + H_{\text{soc}}  
  + H_{\text{Ir-corr}} + H_{\text{O-corr}}.
\end{eqnarray}
In  leading order, the Heisenberg term just vanishes completely.
Ref.~\onlinecite{PhysRevB.78.094403,PhysRevLett.102.017205} further 
considered the splitting among the $t_{2g}$ orbitals and obtained a 
highly anisotropic spin model. For the type-$x$ bond that are in the $yz$ plane, 
the superexchange interaction was found to be~\cite{PhysRevB.78.094403}
\begin{eqnarray}
H_{\text{ex}_1} & = & \sum_{\langle ij \rangle\in x} 
J [- S^x_i S^x_j +  S^y_i S^y_j +  S^z_i S^z_j] 
\label{eq8} 
\end{eqnarray}
and the exchange interaction on other bonds are obtained by simple permutations. 
Jackeli and Khalliulin 
further considered the effect of the Hunds' coupling in Ref.~\onlinecite{PhysRevLett.102.017205} 
and remarkably 
obtained a pure Kitaev interaction for the honeycomb iridate with
\begin{eqnarray}
H_{\text{ex}_2} & = & \sum_{\langle ij \rangle\in x} 
- K  S^x_i S^x_j,
\end{eqnarray}
on the type-$x$ bond, and the interactions on other bonds 
are obtained by simple permutations. 
The anisotropic nature of the superexchange interactions arises 
from the spin-orbit entangled nature and the significant orbital 
content of the Ir$^{4+}$ local moments. 

We close in noting that due to the complication of the extended Hubbard model, 
further theoretical refinements, such as the application of  a Schrieffer-Wolff transformation,
could lead to more complete predictions for the exchange interaction.
Another noteworthy observation
concerns the importance of the intermediate electron configuration
on the interstitial oxygen ions. One could have two holes 
on the same oxygen atom, or one hole on one oxygen and the 
other hole on the other oxygen (see Fig.~\ref{exchange}(b)). 
If one considers these 
intermediate configurations on oxygens, the approximation 
of ``integrating'' out the oxygen $2p$ orbitals/states 
may not be the best approximation, especially when
electron correlations on the $2p$ orbitals are included.

\subsection{Antisymmetric Dzyaloshinskii-Moriya interaction}
\label{Sec:sec2d}

\begin{figure}[t]
	\centering
	\includegraphics[width=0.9\columnwidth]{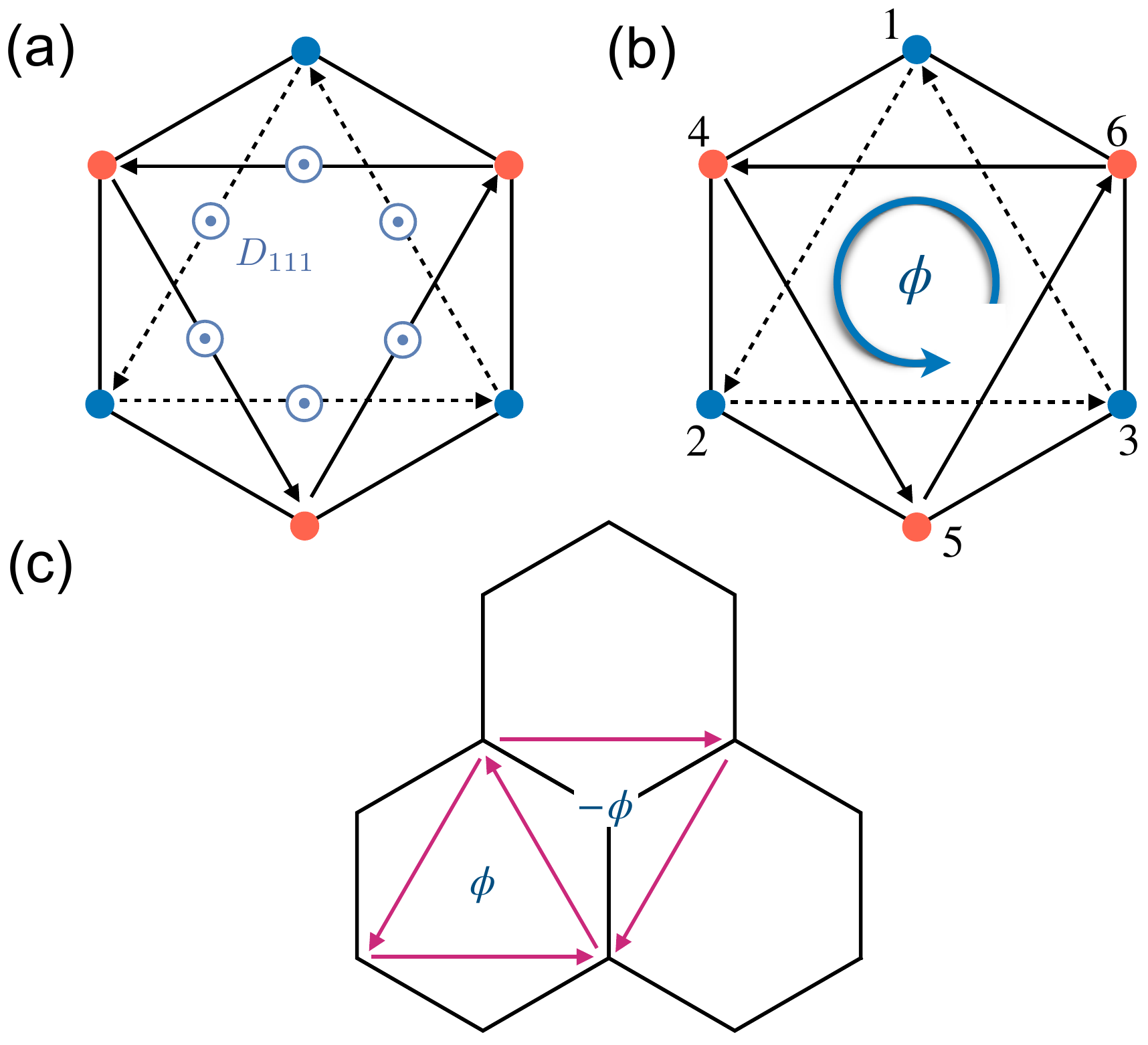}
	\caption{{\bf Dzyaloshinskii-Moriya interactions.}
		(a) Symmetry allowed Dzyaloshinskii-Moriya interactions between 
		second neighbors on the honeycomb lattice, where $D_{111}$ is the $[111]$ component. 
		The arrows specify the order of the cross product 
		$\boldsymbol{S_i}\times\boldsymbol{S_j}$. The two sublattices are labeled by colors.  
		(b) Schematic view 
		of the gauge flux $\phi$ induced by the external magnetic field 
		in the presence of the next-nearest neighbor Dzyaloshinskii-Moriya interaction. 
		(c) The net flux in one unit cell is zero and the space translation symmetry 
		is well preserved.}
	\label{Fig: DMI}
\end{figure}

For $3d$ transition metal compounds with a weak SOC, 
antisymmetric Dzyaloshinskii-Moriya (DM)
interactions~\cite{DZYALOSHINSKY1958241, PhysRev.120.91} are expected
as a higher order perturbation than the pure Heisenberg one
 when the magnetic bonds have no inversion center. 
For magnets with spin-orbit-entangled local moments, the original 
perturbative treatment of SOC~\cite{PhysRev.120.91} to obtain 
the DM interactions no longer applies, but one can 
rely on a symmetry analysis and Moriya's rules to constrain 
the DM interactions. For the two dimensional honeycomb lattice, 
the first neighbor magnetic sites have inversion symmetry, 
thus the first neighbor DM interaction is prohibited. 
However, a second-neighbor DM interaction 
is allowed by symmetry since the second-neighbor magnetic bonds 
have no inversion center. According to Moriya's rules~\cite{PhysRev.120.91}, 
there are components of $\boldsymbol{D_{ij}}$ 
perpendicular to the planes with strength $D_{111}$, as schematically 
depicted in Fig.~\ref{Fig: DMI}(a), and all the in-plane components 
vanish when the honeycomb plane is a mirror plane of the crystal structure. 
Therefore, a representative Dzyaloshinskii-Moriya interaction of the 
honeycomb lattice Mott insulator up to second neighbor has the form, 
\begin{equation}
	H_{\rm DM}=\sum_{ \langle \langle i,j\rangle\rangle }
	\boldsymbol{D}_{ij}\cdot\boldsymbol{S_i}\times\boldsymbol{S_j}.
\label{Eq: DMI}
\end{equation}
For example, it has been estimated~\cite{PhysRevB.93.214431} 
that a large second neighbor Dzyaloshinskii-Moriya term 
${|\boldsymbol{D}_{ij}|>4}$ meV is present for the Kitaev material 
$\alpha$-Li$_2$IrO$_3$, which however is often not considered in the literature.

With these microscopic considerations in place, we note again that our purpose
in the following is not to solve for the ground state of a specific Hamiltonian. 
Instead, we assume that the system stabilizes in the presence of an external 
magnetic field a non-Kitaev spin liquid as suggested by 
numerical studies~\cite{Hickey2019,Lu1809.08247,He1809.09091,PhysRevLett.120.187201} 
and clarify how the elementary spinons in these spin liquids acquire an emergent Lorentz force in the external 
field through the Dzyaloshinskii-Moriya interaction. Due to the Zeeman coupling, 
a moderate magnetic field partially polarizes the spins and generates a finite 
second neighbor scalar spin chirality on the lattice through the 
Dzyaloshinskii-Moriya interaction. 
This in turn induces an internal gauge flux for the spinons, as we will show in the following,
and ultimately give rise to a thermal Hall effect.

\section{Thermal Hall effect for spin liquid with spinon Fermi surface}
\label{Sec:sec3}

As first instance of a non-Kitaev spin liquid we consider the scenario
of a U(1) spin liquid with a spinon Fermi surface. This is motivated
by a recent string of numerical works~\cite{Hickey2019,Lu1809.08247,He1809.09091}
that report strong evidence for the emergence of such a U(1) spin liquid
as an intermediate gapless phase in the antiferromagnetic Kitaev model 
before entering the high-field trivial polarized state.

In more technical terms, the U(1) spin liquid describes a highly entangled quantum 
state with gapless fermionic spinons coupled to a massless U(1) gauge field. 
On a mean-field level, a Hamiltonian  for the neutral spinon 
Fermi surface state can be constructed as 
\begin{equation}
H_{\rm MF}=H_{\rm hop}+H_B,
\label{Eq: Hmf1}
\end{equation}
where $H_{\rm hop}$ contains only spinon hopping operators 
on the honeycomb lattice and 
\begin{equation}
H_B=-\frac{B}{2}\sum_{i,\alpha \beta} f_{i,\alpha}^{\dagger}
(\sigma_x+\sigma_y+\sigma_z)_{\alpha \beta}f_{i,\beta}
\end{equation}
represents the Zeeman coupling to an external magnetic field $B$ 
along the $[111]$ direction, with $f_{i,\alpha}$ ($f_{i,\alpha}^{\dagger}$) 
being the spinon annihilation (creation) operator at site $i$. 
The $[111]$ direction is normal to the honeycomb plane. By studying 
the relation between the relevant projective symmetry 
groups (PSGs)~\cite{PhysRevB.65.165113}, 
three kinds of U(1) spin liquids are obtained~\cite{Lu1809.08247} 
that are connected to the Kitaev $\mathbb{Z}_2$ spin liquid state 
through a continuous phase transition without symmetry breaking. 
Moreover, only 
one of them, labeled as $U_1A_{k=0}$ in Ref.~\onlinecite{Lu1809.08247},
was shown to support robust spinon Fermi surfaces. 
A representative mean-field Hamiltonian for such a state, i.e. a U(1) spin liquid 
with a neutral spinon Fermi surface on the honeycomb lattice, is given in Appendix~\ref{Sec: AppSec1}. 
We will use this mean-field Hamiltonian as our starting point in the following discussion.

\begin{figure}[t]
	\centering
	{\includegraphics[width=3.4 in]{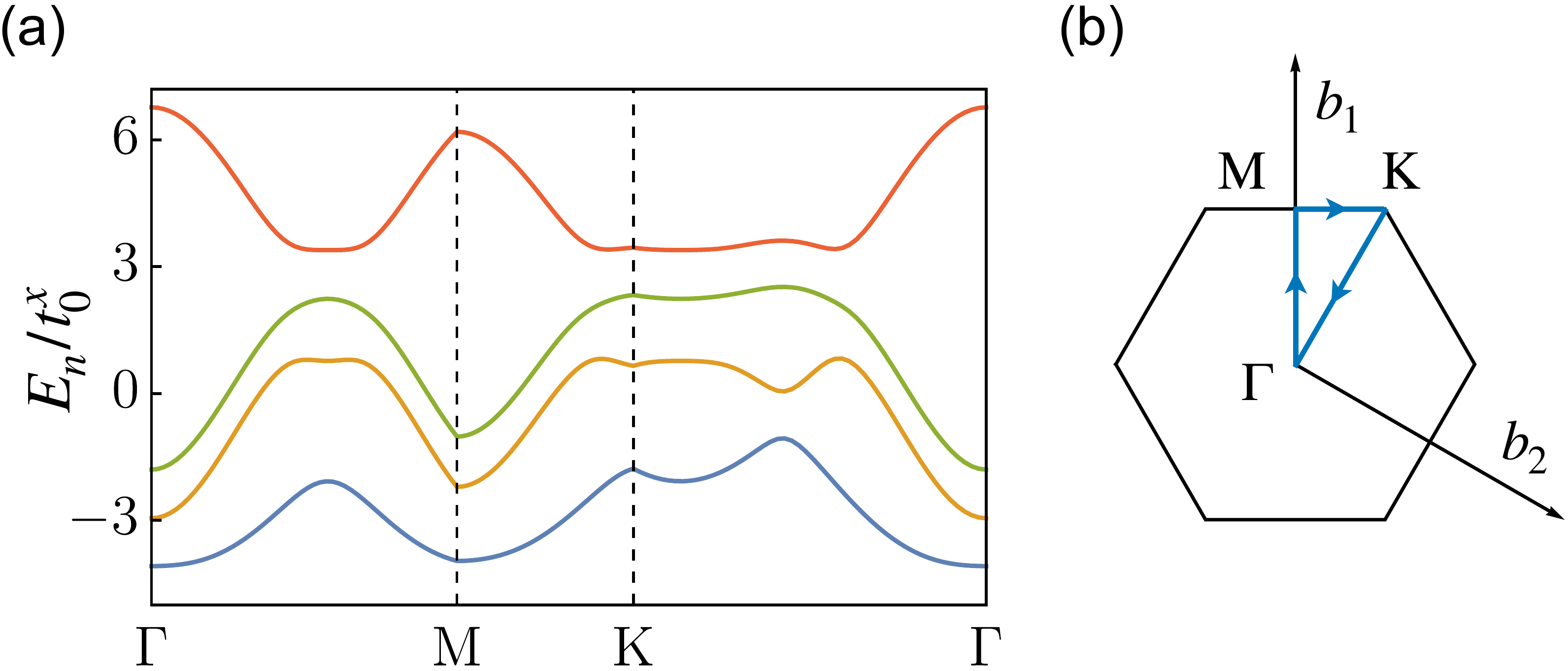}}
	\caption{(a) Representative {\bf spinon dispersions} for the non-zero mean field parameters $(s_3,t_0^x,t_0^y)=(-1,-1,-0.2)$, $(\tilde{s}_0, \tilde{s}_3)=(-0.2,0.2)$ and $(\tilde{t}_0^x,\tilde{t}_0^z,\tilde{t}_3^x,\tilde{t}_3^z)=(-0.2,-0.2,-0.02,-0.02)$ along the high symmetry line. Here the magnetic field is set as ${B=1}$  
	and the induced gauge flux $\phi=\pi/20$. 
	(b) The Brillouin zone of honeycomb lattice 
	with reciprocal lattice vectors $\boldsymbol{b}_1=2\pi (0,2/\sqrt{3})$ 
	and $\boldsymbol{b}_2=2\pi (1,-1/\sqrt{3})$. 
	The arrows indicate the direction of the high symmetry line in (a). }
	\label{Fig: EnergyBand}
\end{figure}

\subsection{Field induced internal flux via Dzyaloshinskii-Moriya interaction}
\label{Sec:sec3b}

For the U(1) spin liquid with spinon Fermi surface in the weak Mott regime, 
by switching on an external magnetic field, a ring exchange interaction 
derived from the Hubbard model can contribute to the thermal Hall 
conductivity~\cite{PhysRevB.51.1922,PhysRevB.73.155115,PhysRevLett.104.066403}. 
It was originally proposed for the well-known triangular lattice organic 
spin liquid candidate $\kappa$-(ET)$_2$Cu$_2$(CN)$_3$, due to its proximity 
to the Mott transition~\cite{PhysRevB.73.155115}. However, since we 
are working in the strong Mott regime, such a mechanism does 
not apply because of the large charge gap. On the other hand, 
as we have mentioned in Sec.~\ref{Sec:sec2d}, the combination 
of the microscopic Dzyaloshinskii-Moriya interaction and Zeeman coupling 
further induces an internal U(1) gauge flux distribution on the honeycomb plane. 

More explicitly, in the U(1) spin liquid phase, 
gauge fluctuations are described by a continuous lattice 
U(1) gauge theory and the internal gauge flux is related 
to the underlying spin chirality 
as~\cite{PhysRevB.39.11413,PhysRevB.46.5621,RevModPhys.78.17}
\begin{equation}
\sin\phi= \frac{1}{2} \boldsymbol{S}_1\cdot\boldsymbol{S}_2\times\boldsymbol{S}_3 \,,
\end{equation}
where $\phi$ is the 
flux defined on the triangular plaquette formed by three 
second neighbor sites of the honeycomb lattice.
Following  previous work by some of us \cite{Chen1901.01522}, 
one can then establish
\begin{equation}
\sin \phi\simeq \lambda D_{111}\chi B/2
\end{equation}
under an external magnetic field $B$ (with $\chi$ 
being the magnetic susceptibility). 
Considering an elementary hexagon as schematically illustrated 
in Fig.~\ref{Fig: DMI}(b), the flux through the triangle formed 
by sites 1, 2 and 3 in the anticlockwise direction corresponds to $\phi$. 
Similarly, the flux through the triangle formed by sites 4, 5 
and 6 in the anticlockwise direction is still $\phi$, i.e., 
the fluxes of the triangles formed by the second neighbor 
bonds in one hexagon are exactly equal for two sublattices. 
However, the flux for the triangle formed by the second 
neighbor bonds of three different hexagons acquires a minus sign 
if adopting the anticlockwise loop convention. That is to say, 
the net flux in one unit cell is zero and the space translation 
symmetry is not destroyed, as shown in Fig.~\ref{Fig: DMI}(c), 
where we only plot the triangles formed by one sublattice for 
simplicity (with an analogue situation for the other sublattice). 
The spinons carry emergent U(1) gauge charges and are minimally 
coupled to the U(1) gauge field, thus the spinons will feel 
such gauge flux as the spinons hop between second neighbor 
sites on the lattice. It is necessary to stress that the first 
neighbor spinon hopping does not pick up any phase since the 
net flux in a unit cell is zero, much like the Haldane model 
for spinless fermions.

\subsection{Reconstructed fermionic spinon bands}
\label{Sec:sec3c}

Physically, as the spinon moves on the lattice, it will experience a 
Lorentz force from the induced internal flux. 
On a semiclassical level, the spinon motion will 
be twisted and reflected, resulting in a spinon thermal Hall effect. 
On a quantum mechanical level, 
this effect can be understood from the spinon Berry curvature, which we explain below.

The internal gauge flux pattern is depicted in Fig.~\ref{Fig: DMI}(b) and (c). 
To capture the flux, we modify the spinon mean-field Hamiltonian by adding 
the U(1) gauge potential to the next-nearest neighbor hopping terms.
This immediately leads to a modified spinon dispersion. Combining the two sublattices 
with the two spin labels, a total of four spinon bands are obtained, 
which are half-filled. As depicted in Fig.~\ref{Fig: EnergyBand}(a), 
the internal U(1) gauge flux reconstructs the spinon bands and there 
still exist Fermi pockets. When the magnetic field exceeds some critical 
value where the pockets vanish, according to Polyakov's argument~\cite{POLYAKOV1977429}, 
the dynamical U(1) gauge field will be confined due to non-perturbative 
instanton events and the system enters a trivial polarized state. 
To describe the thermal Hall effect, we therefore focus on the stable,
deconfined spin liquid regime and further clarify the induced internal 
gauge flux that would contribute to the spinon thermal Hall effect.

\begin{figure}[t] 
\centering
{\includegraphics[width=3.4 in]{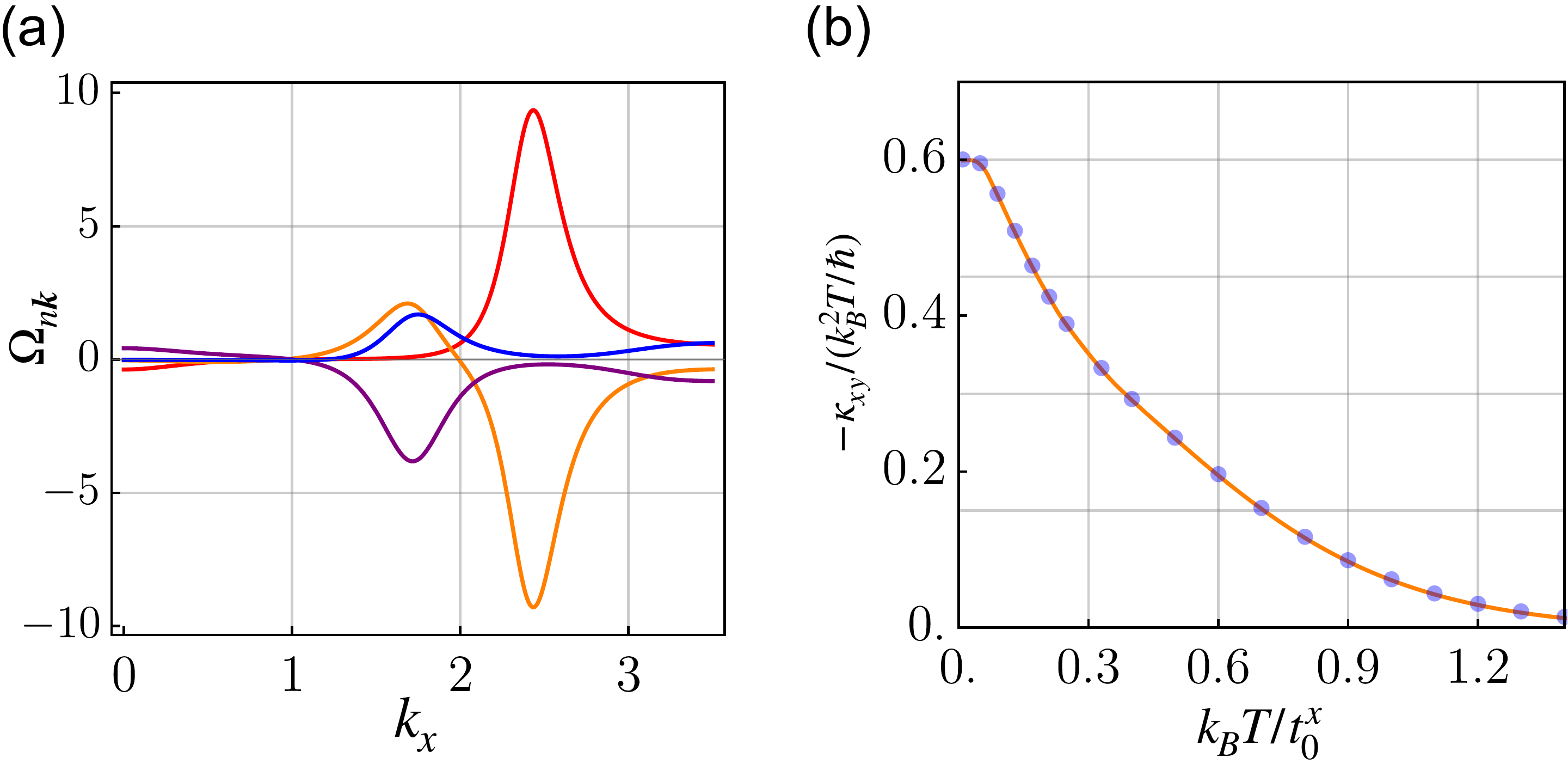}}
\caption{
(a) {\bf Berry curvature} for different energy bands and all 
the parameters are set as in Fig.~\ref{Fig: EnergyBand}. The red 
(orange, purple, blue) line is for the first/lowest (second, third, fourth) 
spinon band. (b) The corresponding evaluation of the {\bf thermal Hall conductivity} 
as a function of temperature.}
\label{Fig: kappa}
\end{figure}

Let us now explicitly demonstrate the finite thermal Hall conductivity 
for the spin liquid in the presence of magnetic field. By the aid of Luttinger's 
pseudogravitational potential~\cite{PhysRev.135.A1505}, 
the thermal Hall conductivity formula for a general noninteracting fermionic system 
with a nonzero chemical potential $\mu$ can be obtained~\cite{PhysRevLett.107.236601} as
\begin{equation}
\label{thermcon}
\kappa_{xy}=-\frac{1}{T}\int d\epsilon(\epsilon-\mu)^2
\frac{\partial f(\epsilon,\mu,T)}{\partial \epsilon}\sigma_{xy}(\epsilon)\,.
\end{equation}	
Here ${f(\epsilon,\mu,T)=1/[e^{\beta(\epsilon-\mu)}+1]}$ is the Fermi-Dirac distribution 
and the derivative of the distribution function 
${\partial f(\epsilon,\mu,T)}/{\partial \epsilon}$ 
indicates that the integral dominates around the Fermi energy. Moreover, 
\begin{equation}
{\sigma_{xy}(\epsilon) 
	= - \frac{1}{\hbar}  
	\sum_{\boldsymbol{k},\xi_{n,\boldsymbol{k}}<\epsilon}\Omega_{n,\boldsymbol{k}}}
\end{equation}
is the zero temperature anomalous Hall coefficient 
for a system with the chemical potential $\epsilon$. 
Here $\Omega_{n\boldsymbol{k}}$ 
is the Berry curvature for the fermions, which is defined as
\begin{eqnarray}
{\Omega_{n\boldsymbol{k}}=-2 {\rm Im}\langle  
	  \frac{\partial u_{n\boldsymbol{k}}}{\partial k_x}|
	  \frac{\partial u_{n\boldsymbol{k}}}{\partial k_y}
	  \rangle}
\end{eqnarray}
with eigenstate $\vert u_{n\boldsymbol{k}} \rangle$ for bands indexed by 
$n$. Eq.~\eqref{thermcon} suggests that the thermal Hall conductivity is 
directly related to the spinon Berry curvature 
in  momentum space and a finite Berry curvature is necessarily required to 
generate $\kappa_{xy}$. We show below that the magnetic field induced internal 
U(1) gauge flux does indeed generate a finite Berry curvature and we can use Eq.~\eqref{thermcon} 
as our basis to calculate the thermal Hall conductivity for the spinon metal in a U(1) 
spin liquid. 
As depicted in Fig.~\ref{Fig: kappa}(a), one can see that the modified mean-field 
Hamiltonian generates non-trivial spinon Berry curvatures for each band due to the 
influence of the induced internal gauge flux. The numerical results for the thermal Hall 
conductivity are presented in Fig.~\ref{Fig: kappa}(b). For a second neighbor 
hopping coefficient ${t_2=0.5t_1}$, we obtain a monotonic temperature dependence 
of $\kappa_{xy}/(k_B^2T/\hbar)$. In the zero temperature limit, it trends to a 
non-zero and non-quantized value. In the finite temperature region,
the thermal Hall conductivity decreases monotonically and finally vanishes  
at high temperatures. The vanishing thermal Hall conductivity in the 
high temperature region originates from the almost equally populated 
spinon bands and the corresponding Berry curvature cancellation.

\subsection{Stability of the U(1) spin liquid}
\label{Sec:sec3d}

\begin{figure}[b]
	\centering
	{\includegraphics[scale=0.2]{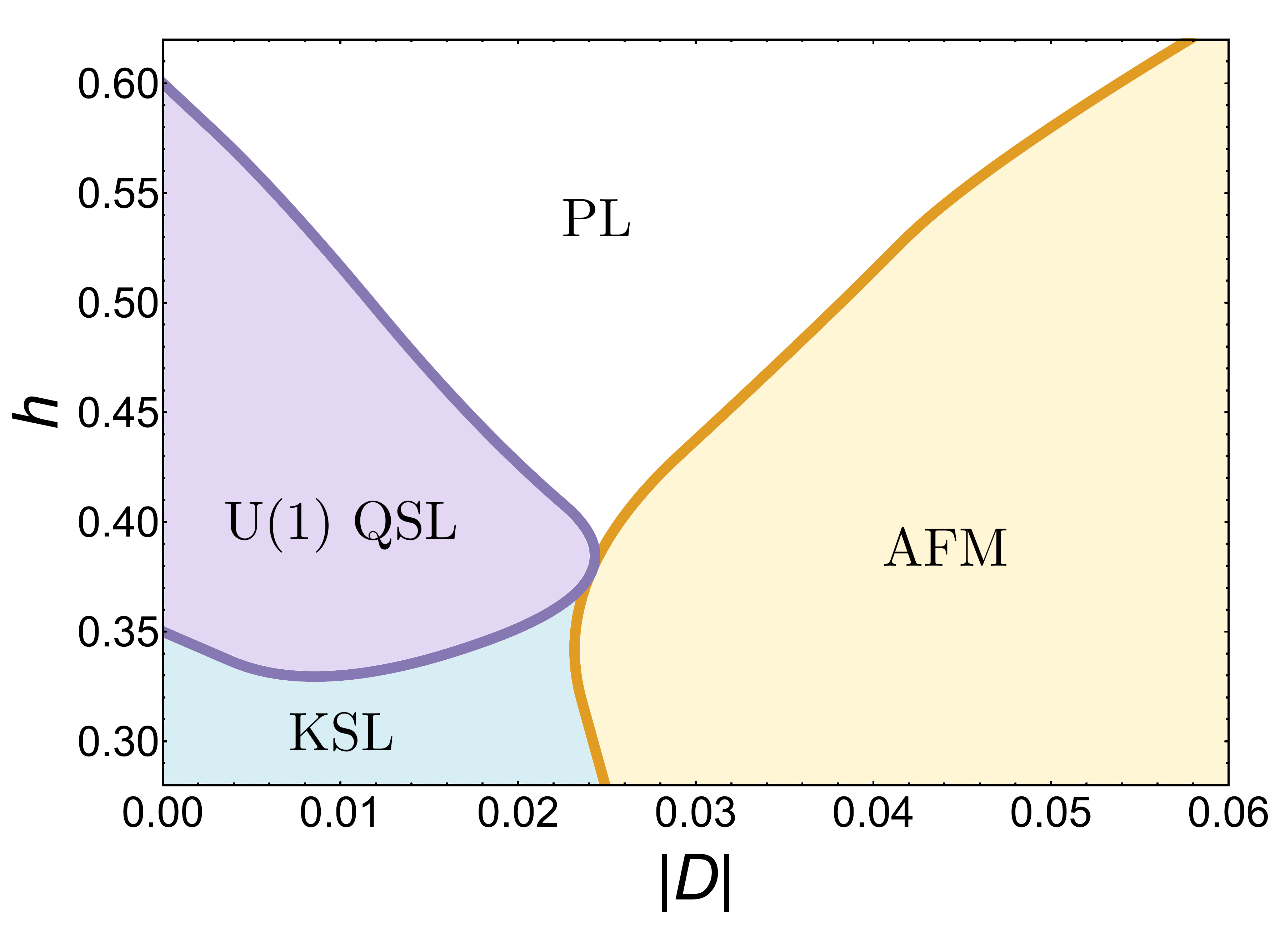}}
	\caption{{\bf Phase diagram for an extended Kitaev model} 
	in the combined presence of a finite Dzyaloshinskii-Moriya interaction, 
	of strength $|\boldsymbol{D}|$, and a finite magnetic field, of magnitude 
	$h$ and oriented close to the $[111]$ direction. The energy unit is in 
	the Kitaev couplinng $K$ of Eq.~\eqref{Eq:EDH}. In the figure, 
	``U(1) QSL'' specifically refers to our spinon Fermi surface U(1) spin liquid,
	``KSL'' refers to the Kitaev spin liquid, ``AFM'' refers to the antiferromagnetic 
	ordered state, and ``PL'' refers to the polarized state. }
	\label{Fig:DM_PD}
\end{figure}

Numerical evidence for a U(1) spin liquid in the Kitaev honeycomb model was 
recently reported for an intermediate magnetic field 
range~\cite{Hickey2019,Lu1809.08247,He1809.09091}. Here, we investigate the 
stability of this U(1) spin liquid to a finite Dzyaloshinskii-Moriya interaction 
using exact diagonalization techniques. For fields close to
the (111) direction the intermediate U(1) spin liquid occurs in a field range 
of $h\sim0.35-0.60K$ (where $h$ is the field magnitude, $h=|\boldsymbol{h}|$). 
We focus on this field range and consider the effects of adding a 
Dzyaloshinskii-Moriya interaction of the form given in Eq.~\eqref{Eq: DMI}. 
The Hamiltonian is thus

\begin{equation}
	H=\sum_{\langle ij \rangle\in \gamma} 
K  S^\gamma_i S^\gamma_j + \sum_{ \langle \langle i,j\rangle\rangle }
	\boldsymbol{D}_{ij}\cdot\boldsymbol{S_i}\times\boldsymbol{S_j} - \sum_i \boldsymbol{h}_i \cdot \boldsymbol{S}_i.
\label{Eq:EDH}
\end{equation}

In Fig.~\ref{Fig:DM_PD} we show the resulting phase diagram, 
with the U(1) spin liquid region stable up to a maximal 
Dzyaloshinskii-Moriya interaction of about $|\boldsymbol{D}|\sim0.025K$. 
We should note however that additional interactions, relevant for real Kitaev materials, 
could further increase or decrease the stability of the U(1) spin liquid 
against the effects of the finite Dzyaloshinskii-Moriya term. In any case, 
the U(1) spin liquid is stable to adding finite, though small, 
Dzyaloshinskii-Moriya interactions. This justifies our starting 
point of U(1) spin liquid even in the presence of Dzyaloshinskii-Moriya interactions.

\section{Thermal Hall effect for Dirac spin liquid}
\label{Sec:sec4}

For particular magnetic field directions on the honeycomb plane, 
a gapless Dirac spin liquid and a gapped Kalmeyer-Laughlin-type~\cite{PhysRevLett.59.2095} 
chiral spin liquid were both numerically obtained in
Ref.~\onlinecite{PhysRevLett.120.187201} for certain parameter regimes
of  the so-called Kitaev-$\Gamma$ model --  a microscopic model with 
additional symmetric off-diagonal $\Gamma$ terms beyond the
Kitaev exchange that has been argued \cite{Rau2016,PhysRevB.93.214431,Winter_2017} 
to be particularly relevant to experimental Kitaev materials.

The gapped chiral spin liquid can be characterized by the net Chern number
of the occupied spinon bands. In addition, note that the ansatz of such a chiral spin liquid 
readily breaks both time-reversal symmetry 
$\mathcal{T}$ and reflection $P$, while their combination $P\mathcal{T}$  
is well preserved. Generically, this leads to a nonvanishing expectation 
value for the chiral order parameter 
${\boldsymbol{S}_i\cdot (\boldsymbol{S}_j\times\boldsymbol{S}_k)}$, 
where $i, j, k$ are three nearby sites. The chiral spin liquid is 
effectively described by the Chern-Simons theory with semion 
topological order, especially, this state has chiral edge modes 
and would show an integer-quantized thermal Hall effect. Thus 
we are not going to further discuss the influence of the induced 
internal gauge flux on this state due to the Chern-Simons term in 
the theory for gauge fluctuations.

\begin{figure}[b]
\centering
{\includegraphics[width=3.4 in]{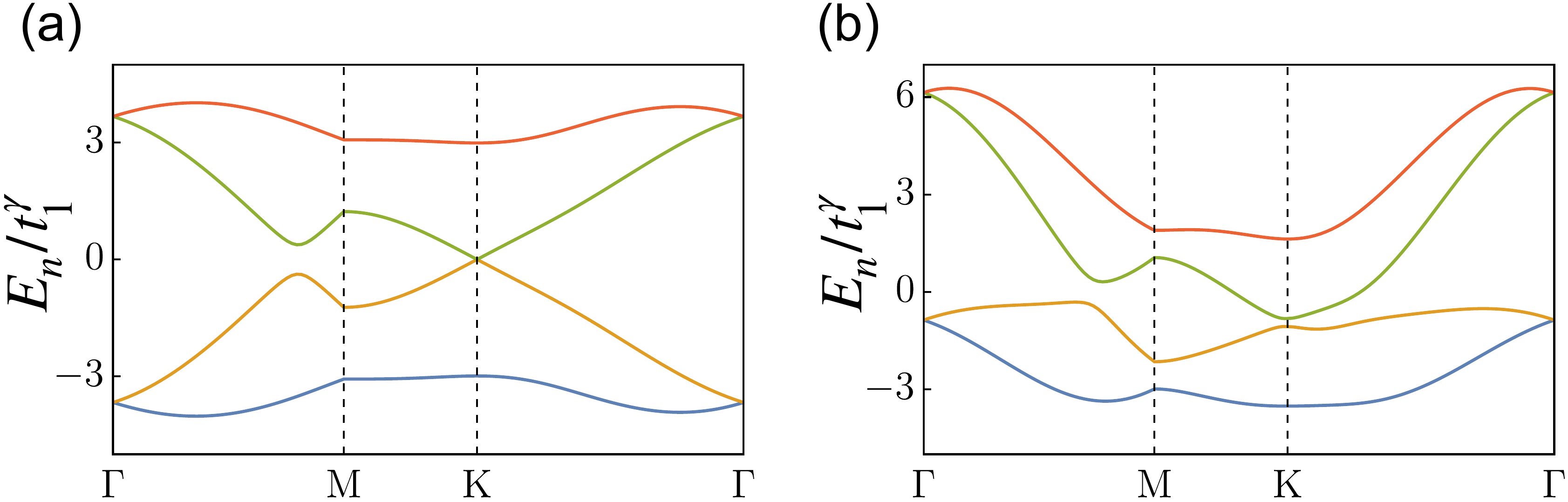}}
\caption{(a) {\bf Spinon dispersion for the Dirac spin liquid}. 
(b) The induced flux for the second neighbor hopping terms 
reconstructs the spinon bands and the resulting state is a spinon 
Fermi surface spin liquid. There is a Fermi pocket around the 
${\rm K}$ point of the Brillouin zone.}
\label{Fig: EnergyBand2}
\end{figure}

Here we consider the situation where the system stabilizes and 
stays in a gapless Dirac spin liquid state. Such a Dirac spin liquid is a deconfined 
state with Dirac band touchings at the Fermi level and its 
low-energy effective theory is described by the Dirac equation. 
Usually, a Dirac spin liquid  has no thermal Hall effect associated with it. 
A representative spinon dispersion for the Dirac spin liquid 
realized in the Kitaev-$\Gamma$ model for the honeycomb lattice 
is depicted in Fig.~\ref{Fig: EnergyBand2}(a), where we have 
adopted the spinon mean-field Hamiltonian constructed in 
Ref.~\onlinecite{PhysRevLett.120.187201} 
(see Appendix \ref{Sec: AppSec2} for details). 
One can see that, at the Fermi level, there is a Dirac band 
touching at the ${\rm K}$ point of the Brillouin zone. 
We assume that this deconfined spin liquid state is stabilized 
in a finite region of the phase diagram and the presence of 
the second neighbor Dzyaloshinskii-Moriya interaction would not destroy it.

\begin{figure}[t] 
	\centering
	{\includegraphics[width=7cm]{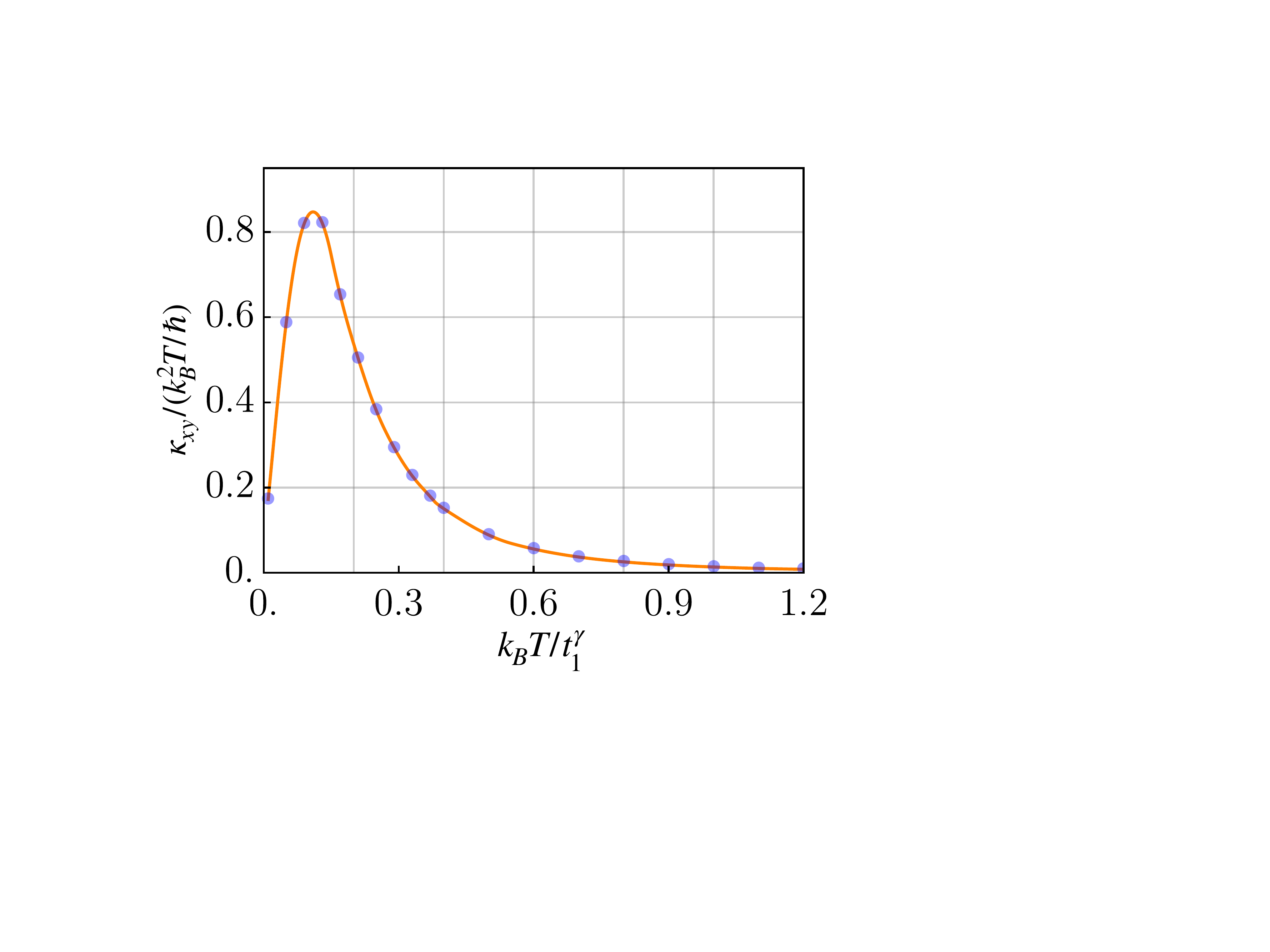}}
	\caption{The {\bf thermal Hall 
	conductivity for Dirac spin liquid} as a function of 
	temperature at the gauge flux ${\phi=\pi/10}$, 
	here $t_2$ is set as 0.4 $t_1^{\gamma}$.}
	\label{Fig: kappa2}
\end{figure}

As in the spinon Fermi surface U(1) spin liquid case, the gauge fluctuations 
of the Dirac spin liquid are described by a U(1) gauge theory, 
thus the external magnetic field also induces an internal gauge 
flux for the second neighbor spinon hopping channels through the 
second neighbor Dzyaloshinskii-Moriya interaction and leads to 
a spinon thermal Hall effect. Such flux will reconstruct the 
spinon bands and the resulting state is a spinon Fermi surface 
spin liquid with Fermi pocket around the ${\rm K}$ point, 
as shown in Fig.~\ref{Fig: EnergyBand2}(b). Although the Dirac 
band touching is eliminated, when we consider the influence of the gauge flux, 
the system is still in a deconfined phase since the matter field 
is also gapless and the gap between the second and third bands 
is not relevant. Following a similar procedure as for the calculation 
in Sec.~\ref{Sec:sec3}, in Fig.~\ref{Fig: kappa2} we plot the 
temperature dependence of the thermal Hall conductivity for this state. 
In contrast to Fig.~\ref{Fig: kappa}(b), the ratio of thermal Hall 
conductivity and temperature for this state increases rapidly with 
temperature and then decreases gradually after reaching a maximum 
in a finite temperature region. Such a different temperature 
dependence originates from the special spinon dispersion and the 
corresponding spinon Berry curvature of this state. The 
vanishing of the thermal Hall conductivity in the high 
temperature region can again be explained by the  
Berry curvature cancellation of different spinon bands.

\section{Discussion}
\label{Sec:sec5}

While the original motivation for the exploration of the growing family of Kitaev materials~\cite{WitczakKrempa2014,Rau2016,Trebst2017} 
might have been to discover an experimental realization of the Kitaev spin liquid~\cite{KITAEV20062}, 
i.e.~a non-Abelian chiral spin liquid with a gapless Majorana edge current, 
it is becoming increasingly clear that these materials might also harbor 
other types of spin liquids~\cite{Hickey2019,Lu1809.08247,He1809.09091,
PhysRevLett.120.187201,PhysRevB.99.205119}. Theoretical investigations
suggest that this is particularly true when considering field-induced phases, 
for which the emergence of a U(1) spin liquid with spinon Fermi 
surface~\cite{Hickey2019,Lu1809.08247,He1809.09091}, 
a Dirac spin liquid~\cite{PhysRevLett.120.187201} 
and Abelian chiral spin liquids~\cite{PhysRevLett.120.187201} 
have been proposed. In the absence of magnetic fields, 
additional types of $\mathbb{Z}_2$ spin liquids, beyond the ones known from the bare Kitaev model, 
have been proposed 
to arise from strong spin-orbit coupling~\cite{PhysRevB.99.205119}. 
However, narrowing in on a precise theoretical prediction starting from 
{\em ab initio} modeling~\cite{Rau2016,PhysRevB.93.214431}
for a given Kitaev material remains a formidable challenge, 
not least because of the myriad of additional couplings that 
are at play in these materials beyond the pure Kitaev exchange~\cite{Winter_2017}, 
including isotropic Heisenberg interactions of varying range 
and spin-orbit induced off-diagonal spin exchanges or Dzyaloshinskii-Moriya interactions. 
This is particularly so for non-honeycomb based iridates 
with a three-dimensional structure, as there more complex 
lattice geometries allow for even more types of symmetry-allowed magnetic interactions.

It might thus be a better strategy to instead start from the potential 
spin liquid states and to single out experimental signatures that allow 
to distinguish between these different non-magnetic states. 
As we have argued in this manuscript, the observation of a thermal Hall effect 
is precisely such a measure. In the absence of magnetic orders, 
it allows one to single out the nature of the potential 
candidate spin liquids: a finite, but non-quantized, 
thermal Hall effect is indicative of a {\em gapless} spin liquid,
either with a spinon Fermi surface or a Dirac spectrum, 
while a quantized thermal Hall effect exposes
a {\em gapped} chiral spin liquid whose Abelian versus 
non-Abelian character is reflected in the integer versus
half-integer quantization of the edge modes.

\subsection*{Application to H$_3$LiIr$_2$O$_6$}

Amongst the honeycomb Kitaev materials, the recently synthesized 
H$_3$LiIr$_2$O$_6$~\cite{Takagi} stands out as the only material 
that remains disordered down to the lowest measured temperature.
As such it might be the best candidate material to date to exhibit 
a (gapless) Kitaev spin liquid even in the absence of magnetic field. 
Experimentally, the system exhibits a constant susceptibility
and sub-linear power law heat capacity at low temperatures~\cite{Takagi}. 
The observation of constant magnetic susceptibility might not be unexpected
in light of the fact that the Ir $5d$ electrons are subject to strong SOC, 
which in turn renders the notion that the magnetization remains a good quantum 
number down to zero temperature invalid as discussed early on~\cite{PhysRevB.78.094403}.
To rationalize the unusual scaling behavior of the specific heat, which must be rooted in a divergent 
low-energy density of states, several explanations have been put forward that start from the gapless
Kitaev spin liquid and consider the effect of additional perturbations, 
such as a residual interlayer coupling~\cite{PhysRevB.97.115159}
or disorder effects~\cite{PhysRevLett.121.197203,PhysRevLett.121.247202,PhysRevLett.122.047202,Kimchi}.
Indeed {\em ab initio} calculations indicate that the local ${j=1/2}$ 
moments experience a significant amount of quenched bond disorder 
arising from structural disorder of the H ions~\cite{PhysRevLett.122.047202,PhysRevLett.121.197203}.
On a phenomenological level, the formation of a disorder-induced 
random singlet phase has been put forward~\cite{Kimchi}
\footnote{
The formation of such a random singlet phase has been conceptualized as an extension
of a one-dimensional random singlet phase to high dimensions with spin-orbit anisotropy
\cite{Kimchi}.
Note, however, that there are some key differences from the one-dimensional scenario. 
For a random antiferromagnetic spin-1/2 chain, the random singlet phase can be obtained 
in a rather elegant calculation performing a real space renormalization group and 
master equation flow~\cite{PhysRevLett.69.534,PhysRevB.50.3799,PhysRevB.51.6411}, 
which is asymptotically exact. However, this asymptotic exactness does not 
easily generalize as has been recognized for other one-dimensional problems, 
such as disordered boson chains~\cite{PhysRevLett.100.170402}. In higher 
spatial dimensions, the higher connectivity of every site and the sign of 
the interactions may further complicate the problem, and the formulation 
of a strong-disorder renormalization group approach for two-dimensional 
systems~\cite{PhysRevB.61.1160,PhysRevB.65.064206,PhysRevB.85.224201}
remains a formidable challenge. As such, endowing the proposal for a 
two-dimensional random singlet phase with a more substantial theoretical 
footing may remain an open and interesting subject at this stage.}. 
Considering these different scenarios, one naturally arrives at 
the question of how one can distinguish the different potential origins 
for the apparent non-magnetic behavior of H$_3$LiIr$_2$O$_6$ -- disorder effects
versus the formation of a random singlet phase or the emergence of 
a spin liquid -- in experimental probes.

As we have argued above, performing thermal Hall measurements 
for small external magnetic fields on the single crystalline 
samples of H$_3$LiIr$_2$O$_6$, which have recently become available, 
would provide distinct insight -- with the observation of a thermal 
Hall signature being direct evidence for a spin liquid scenario.  
If the disorder is in fact weak and the system is indeed in a gapless 
Kitaev spin liquid, the field will induce a transition to a fully gapped Kitaev spin liquid
that may overcome the disorder effect and show a half-quantized thermal Hall effect,
similar to what has been observed for RuCl$_3$. If one instead observes a finite, 
but non-quantized thermal Hall effect this would count as evidence 
for the formation of a non-Kitaev spin liquid, either with a spinon Fermi surface 
or a Dirac cone spectrum.

\subsection*{Conclusions}

With our present study we have completed an analysis of the thermal Hall signatures
of various Kitaev and non-Kitaev spin liquids that have been discussed as candidate phases
in the context of Kitaev materials in an external magnetic field. 
Complementing earlier studies on the conventional Kitaev spin liquid, a chiral spin liquid 
with a topological Majorana fermion band structure, we have considered, in detail, the
spinon thermal Hall effect arising for various non-Kitaev spin liquids, in particular 
gapless U(1) spin liquids with a spinon Fermi surface~\cite{Hickey2019,Lu1809.08247,He1809.09091}, 
Dirac spin liquids, and variants of Abelian chiral spin liquids~\cite{PhysRevLett.120.187201}.

The mechanism for the appearance of a finite thermal Hall effect in the case of 
U(1) spin liquids, namely the interplay of Dzyaloshinskii-Moriya interactions 
and Zeeman coupling, also results in a specific angular dependence of the sign 
of the measured thermal Hall conductivity. Specifically, the sign of the Hall 
conductivity is fixed by the sign of the dot product between the 
Dzyaloshinskii-Moriya vector with the external magnetic field, 
i.e.\ whether they are parallel or anti-parallel. In the case of 
$\alpha$-RuCl$_3$, where the Dzyaloshinskii-Moriya vector 
is perpendicular to the honeycomb planes, this means that the 
sign is simply given by the sign of the out-of-plane component 
of the field, $\text{sgn}(B_{111})$.

It is important to note that a finite thermal Hall effect 
can also arise in a magnetically {\em ordered} state, 
where it arises from a non-trivial Berry curvature of the 
elementary magnon bands~\cite{PhysRevB.98.060412}.
As such, it is of paramount importance to first establish 
whether a given material exhibits any ordering tendencies 
in the temperature and magnetic field regime of interest, 
in order to distinguish whether a possible thermal Hall
signature arises from conventional magnons or in fact spinons, 
which would be a strong indication for the fractionalization 
inherently connected to quantum spin liquid formation.

Besides the thermal Hall effect, there have been several 
recent theoretical works that attempt to understand the 
magnetic field effect on the spinon Fermi surface state. 
For the case of strong Mott insulators, where only the Zeeman coupling
needs to be considered, Ref.~\onlinecite{PhysRevB.96.075105} established
the spectral evolution of the spinon continuum and a spectral 
crossing in the continuum within the free spinon mean-field theory.
More recently, Ref.~\onlinecite{Balents2019} included the effect of spinon-gauge coupling
as well as short-range spinon interactions to
predict the existence of a new collective mode dubbed ``spinon wave mode''.
For the case of weak Mott insulators, recent work~\cite{Inti2019} 
considered the orbital coupling to the magnetic field~\cite{PhysRevB.73.155115} 
rather than a Zeeman coupling
to estimate the cyclotron resonance of the spinon Fermi surface state.

\section{Acknowledgments}

Y.H.G. and G.C. thank Jia-Wei Mei and Shiyan Li for an update on the experimental analysis
of RuCl$_3$, and Xuefeng Sun from USTC for clarifying the experimental setups. 
This work is supported by the Ministry of Science and Technology 
of China with Grant No.2016YFA0301001, 2016YFA0300500, 2018YFGH000095 
and by the Deutsche Forschungsgemeinschaft (DFG, German Research Foundation) -- 
project numbers 277101999 and 277146847 --
TRR 183 (project B01) and CRC 1238 (project C02).

\begin{appendix} 
	
\section{Spinon Fermi surface mean-field Hamiltonian} 
\label{Sec: AppSec1}

When we discuss the spinon Fermi surface spin liquid realized in this system under an intermediate magnetic field, we mainly follow the  analysis in Ref.~\onlinecite{Lu1809.08247}, and use the results and conclusions therein. In the numerical calculation, the physical spin model that was used is the original Kitaev model with magnetic field and exchange anisotropy. For the convenience of the presentation in the work, we also list one of their mean-field Hamiltonians on which we focus here. In the momentum space, the spinon mean-field Hamiltonian for the $U_1A_{k=0}$ state with a neutral spinon Fermi surface has the following form
\begin{align}
H_{\rm NFS} = \sum_{\boldsymbol{k}} \Psi_{\boldsymbol{k}}^\dagger(h_{0\boldsymbol{k}} + h_{1\boldsymbol{k}} + h_{2\boldsymbol{k}}) \Psi_{\boldsymbol{k}},
\label{Eq: NSFH}
\end{align}
with the {\bf k}-space	basis $\Psi_{\boldsymbol{k}} = (a_{\boldsymbol{k}\uparrow}, a_{\boldsymbol{k}\downarrow}, b_{\boldsymbol{k}\uparrow}, b_{\boldsymbol{k}\downarrow})^T$, and $a_{\bf k}$ and $b_{\bf k}$ are for A and B sublattices, respectively. Moreover, the numbers 0,1,2 denote onsite, nearest neighbor, and next-nearest neighbor terms, respectively, as defined below.

The onsite terms are given by
\begin{align}
	h_{0} = -\mu \tau_0\sigma_0 - \frac{B}{8} \tau_0 (\sigma_x+\sigma_y + \sigma_z),
\end{align}
where $\mu$ is the spinon chemical potential and would be self-consistently calculated by the Hilbert space constraint 
\begin{align}
	\sum_{\sigma}f_{i\sigma}^{\dagger}f_{i\sigma}=1.
\end{align}
Nearest neighbor terms are
\begin{align}
	h_{1\boldsymbol{k}} &= - \begin{pmatrix}
	0 & D_{\boldsymbol{k}} \\ D_{\boldsymbol{k}}^\dagger & 0
	\end{pmatrix}\\ \nonumber
	D_{\boldsymbol{k}} &= (s_3\sigma_0 + t^x_0\sigma_x + t^y_0\sigma_y + t^y_0\sigma_z) e^{-ik_1}\\ \nonumber
	&  + (s_3\sigma_0 + t^y_0\sigma_x + t^x_0\sigma_y + t^y_0\sigma_z) e^{-ik_2}\\
	&  + (s_3\sigma_0 + t^y_0\sigma_x + t^y_0\sigma_y + t^x_0\sigma_z)
\end{align}
and next-nearest neighbor terms are
\begin{align}
	h_{2\boldsymbol{k}} =&\, -\begin{pmatrix}
	A_{\boldsymbol{k}} & 0 \\ 0 & B_{\boldsymbol{k}}
	\end{pmatrix},\\ \nonumber
	A_{\boldsymbol{k}} =&\, 2 (\tilde{s}_0\sigma_0 + \tilde{t}^x_3\sigma_x + \tilde{t}^x_3 \sigma_y + \tilde{t}^z_3\sigma_z) \sin(k_1-k_2)\\ \nonumber
	& + 2(\tilde{s}_3\sigma_0 + \tilde{t}^x_0\sigma_x + \tilde{t}^x_0\sigma_y + \tilde{t}^z_0\sigma_z) \cos(k_1-k_2)\\
	\nonumber
	& + 2 (\tilde{s}_0 \sigma_0 + \tilde{t}^z_3 \sigma_x + \tilde{t}^x_3 \sigma_y + \tilde{t}^x_3 \sigma_z) \sin (-k_2)\\ \nonumber
	& + 2 (\tilde{s}_3 \sigma_0 + \tilde{t}^z_0\sigma_x + \tilde{t}^x_0 \sigma_y + \tilde{t}^x_0\sigma_z) \cos (-k_2)\\ \nonumber
	& + 2 (\tilde{s}_0\sigma_0 + \tilde{t}^x_3 \sigma_x + \tilde{t}^z_3 \sigma_y + \tilde{t}^x_3\sigma_z) \sin (k_1)\\
	& + 2 (\tilde{s}_3+ \tilde{t}^x_0\sigma_x + \tilde{t}^z_0 \sigma_y + \tilde{t}^x_0 \sigma_3) \cos (k_1),\\ \nonumber
	B_{\boldsymbol{k}} =&\, 2(\tilde{s}_0\sigma_0 + \tilde{t}^x_3\sigma_x + \tilde{t}^z_3\sigma_y + \tilde{t}^x_3\sigma_z) \sin(-k_1)\\ \nonumber
	& + 2(\tilde{s}_3\sigma_0 + \tilde{t}^x_0\sigma_x + \tilde{t}^z_0\sigma_y + \tilde{t}^x_0\sigma_z) \cos(-k_1)\\ \nonumber
	& + 2(\tilde{s}_0\sigma_0 + \tilde{t}^x_3\sigma_x + \tilde{t}^x_3\sigma_y + \tilde{t}^z_3\sigma_z) \sin[-(k_1-k_2)]\\ \nonumber
	& + 2(\tilde{s}_3 \sigma_0 + \tilde{t}^x_0\sigma_x + \tilde{t}^x_0 \sigma_y + \tilde{t}^z_0\sigma_z) \cos[-(k_1-k_2)]\\ \nonumber
	& + 2(\tilde{s}_0 \sigma_0 + \tilde{t}^z_3\sigma_x + \tilde{t}^x_3\sigma_y + \tilde{t}^x_3\sigma_z) \sin(k_2) \\
	& + 2(\tilde{s}_3 \sigma_0 + \tilde{t}^z_0\sigma_x + \tilde{t}^x_0 \sigma_y + \tilde{t}^x_0\sigma_z) \cos(k_2).
\end{align}

In the above scenario, we have labeled the two dimensional momentum 
by ${{\bf k}=k_1\boldsymbol{b}_1+k_2\boldsymbol{b}_2}$, where 
$\boldsymbol{b}_1=2\pi (0,2/\sqrt{3})$ and 
${\boldsymbol{b}_2=2\pi (1,-1/\sqrt{3})}$
are reciprocal lattice vectors associated with Bravais lattice vectors 
$\boldsymbol{a}_1,\boldsymbol{a}_2$, thus we have
\begin{eqnarray}
	k_1 &=& \frac{\sqrt3 k_x + 3k_y}{2},\\
	k_2 &=& \frac{-\sqrt3 k_x + 3k_y}{2}.
\end{eqnarray}
This mean-field spinon Hamiltonian Eq.~\ref{Eq: NSFH} is our basis 
to further discuss the thermal Hall effect in the spinon Fermi 
surface spin liquid in Sec.~\ref{Sec:sec3} of the main text.

\section{Dirac spin liquid mean-field Hamiltonian} 
\label{Sec: AppSec2}

When we consider the Dirac spin liquid case, we instead follow the analysis in Ref.~\onlinecite{PhysRevLett.120.187201}. In their work, with respecting the lattice symmetry of the Kitaev-${\rm \Gamma}$ model, the mean-field Hamiltonian constructed  for the Dirac spin liquid is given by
\begin{align}
	H_{\rm Dirac}=&\sum_{\langle ij \rangle \in \alpha\beta(\gamma)} [C_i^\dag (t_1^\gamma 
	R_{\alpha\beta} - it_0^\gamma + t_2^\gamma \sigma_\gamma) C_j + {\rm H.c.}] \nonumber 
	\\&+g \mu_B \sum_i C_i^\dag ({\textstyle \frac12} \pmb B \! 
	\cdot \! {\pmb \sigma} + \lambda) C_i + H_0,
\end{align}
where ${R_{\alpha\beta} = {-i\over\sqrt2}(\sigma_\alpha + \sigma_\beta)}$, 
${t_1^\gamma= - {\textstyle \frac12} |K| \langle C_i^\dag R_{\alpha\beta} C_j \rangle^*}$, ${t_{0,2}^\gamma = - {\textstyle \frac{1}{8}}(\Gamma - |K|) [\langle C_i^\dag \sigma_\alpha R_{\alpha\beta} C_j \rangle^* \pm \langle C_i^\dag \sigma_\beta R_{\alpha\beta} C_j \rangle^*]}$, $\lambda$ is a Lagrange multiplier corresponding to the average particle-number constraint to ensure the proper physical Hilbert space, functioning as a chemical potential, and $H_0$ is a constant. The $t_1^\gamma$ and $t_2^\gamma$ terms are analogous to the Rashba SOC of electrons. Rigorously speaking, all of these parameters should be determined by variational Monte Carlo calculations in which the local constraint is enforced exactly, as numerically studied by Liu and Normand \cite{PhysRevLett.120.187201}. Here for convenience, we simply take some values of them to illustrate our idea of spinon thermal Hall effect in the U(1) spin liquid, but notice that when we choose these parameters, we always guarantee that the spinon (matter) field is gapless so that the system is in a deconfined phase.

\section{Duality properties of symmetric spin models}
\label{Sec: sec2b}

\begin{figure}[t]
	\centering
	\includegraphics[width=4.3cm]{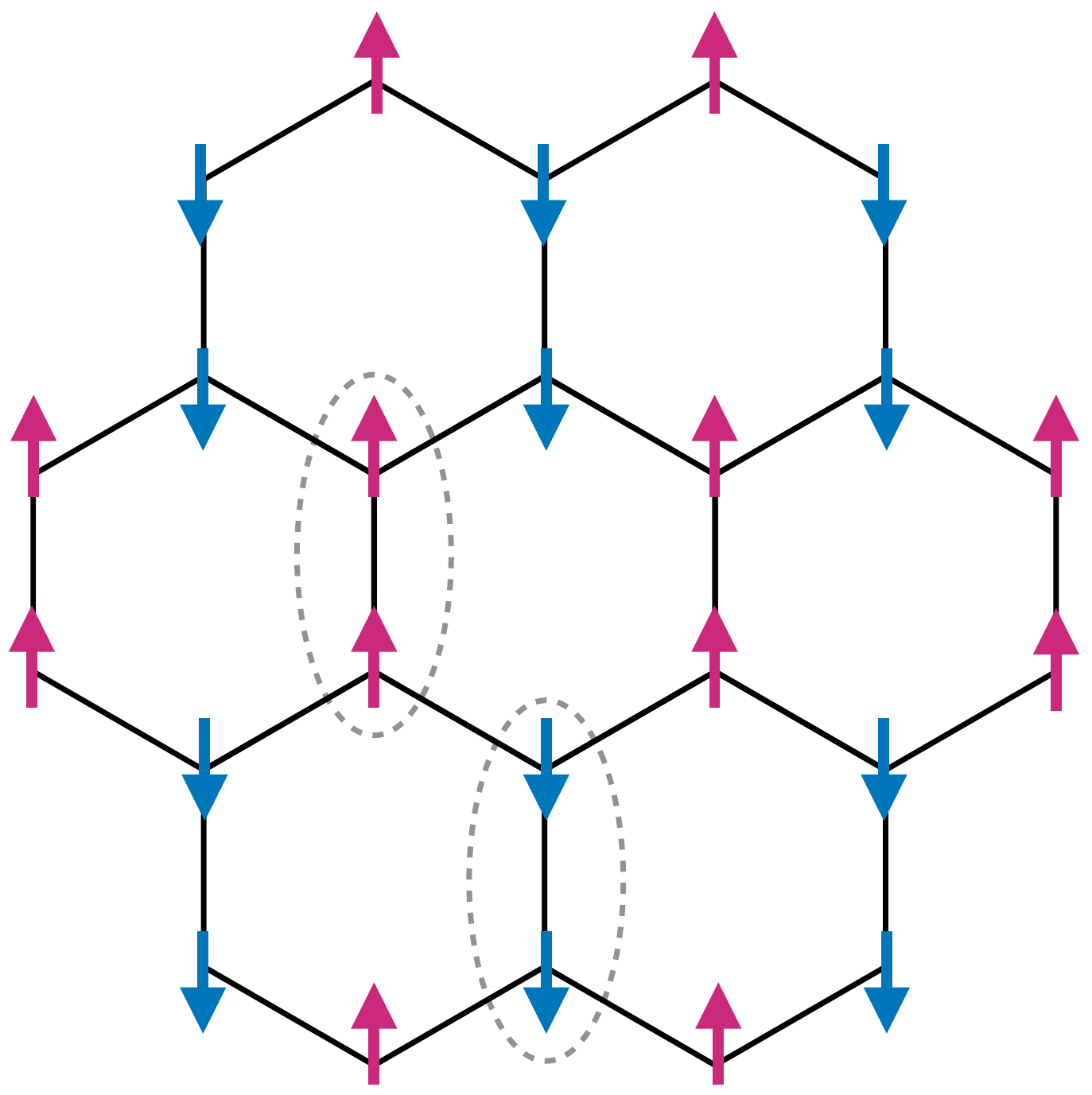}
	\caption{{\bf Stripy antiferromagnetic state} on the honeycomb lattice. Each stripye is composed of the two-site clusters (circled by the dashed gray ellipse) that reside on a given line.}
	\label{Fig: stripyAFM}
\end{figure}

In spatial dimension higher than one, exactly solvable 
quantum Hamiltonians are rather scarce. Very interestingly, 
the anisotropic Hamiltonian Eq.~\eqref{eq8} for the honeycomb 
lattice was found to be exactly solvable since it can be mapped to
 a simple Heisenberg model on all bonds simultaneously, 
 with a hidden ferromagnet exposed by the site-dependent 
 spin rotation that quadruples the original unit cell. 
 This mapping has been known as the four-sublattice spin 
 rotation trick after a work~\cite{PhysRevLett.89.167201} 
 for $t_{2g}$ orbitals in a cubic environment, whose general 
 structure was later elucidated and 
 referred to as Klein duality in Ref.~\onlinecite{PhysRevB.89.014414}. 
 The site-dependent $\pi$ rotations of the four-sublattice 
 spin transformation connect to the Kitaev exchange through 
 the multiplication rules of the Klein four group.

To be concrete, for the honeycomb lattice, we consider the 
rotated spin operators $\tilde{\boldsymbol{S}}$ where 
$\tilde{\boldsymbol{S}}=\boldsymbol{S}$ for one sublattice and,  
depending on the particular sublattice they belong to, $\tilde{\boldsymbol{S}}$ on the remaining three sublattices differ from the original $\boldsymbol{S}$ by the sign of two appropriate components~\cite{PhysRevLett.105.027204}. Written in the rotated basis, Eq.~\eqref{eq8} reads
\begin{equation}
H_{\rm ex_{1}}=\sum_{ \langle i,j\rangle \in x}-J\tilde{\boldsymbol{S}}_i \cdot\tilde{\boldsymbol{S}}_j
\label{Eq: kdH}
\end{equation}
with a ferromagnetic interaction. It is straightforward to obtain from Eq.~\eqref{Eq: kdH} that the exact ground state of Eq.~\eqref{eq8} is a fully polarized ferromagnetic state in the rotated basis. After applying the rotation defined by the Klein duality on this magnetic order and mapping it back to the original spin basis, the resulting order is depicted in Fig.~\ref{Fig: stripyAFM}, which corresponds to a stripy collinear antiferromagnetic pattern of the original magnetic moments. Each stripy is composed of the two-site clusters (circled by the dashed gray ellipse in Fig.~\ref{Fig: stripyAFM}) that reside on a given line. Despite belonging to an antiferromagnetic type, this stripy order is fluctuation-free and would show a fully saturated antiferromagnetic order parameter~\cite{PhysRevLett.105.027204}.

Moreover, to make use of the Klein duality, the model should usually be considered to be just the pure nearest-neighbor Kitaev-Heisenberg model. The duality properties generally break down when the Dzyaloshinskii-Moriya interaction and/or other further-neighbor interactions are included.

\section{Electron-hole doping asymmetry}
\label{Sec:sec2c}

Since the proposal of the Kitaev model and Kitaev spin liquid in real materials,
there has been one direction of effort in the field that tries to study the 
effect of dopings in the relevant materials~\cite{PhysRevB.86.085145}. 
This is certainly a natural 
direction of thinking, on both the model level and the experimental level. 
One strong motivation comes from the cuprate superconducters that were often
viewed as a doped Mott insulating spin liquid. The pairing already occurs 
in the spin liquid regime, and condensed doping holes generate superconductivity. 
If one of the honeycomb Kitaev materials realizes the Kitaev spin liquid, doping it 
would probably generate topological superconductivity. This statement, however,
ignores the detailed evolution of spin-orbital structure of the system under
doping. There is an electron-hole doping asymmetry for such materials. 
If doping happens on the transition metal ions, the electron doping would create a 
$d^6$ electron configuration that has no spin or orbital structure.
In contrast, hole doping would create a $d^4$ electron configuration that 
has a reconstructed spin-orbit structure~\cite{PhysRevB.84.094420}. 
The local energy level would be 
a total ${J=0}$ ground state, ${J=1}$ triplet excited states and ${J=2}$ quintuplets~\cite{PhysRevB.84.094420}. 
This would create a big difference between the electron doping and the hole
doping. One may compare with cuprates where an electron-hole doping asymmetry 
also occurs. There, electrons are doped on the Cu sites, while holes are doped 
on the O sites~\cite{PhysRevLett.95.137001}. 
This is due to the charge transfer nature of the insulating phase.

This electron-hole doping asymmetry and the reconstructed spin-orbital structure 
occur quite generically  
in the strong spin-orbit-coupled correlated materials with rather different 
electron configurations, beyond iridates or honeycomb Kitaev materials. An earlier work 
that considered to dope $d^4$ Mott 
insulators with excitonic magnetism has noticed that doping constructs the 
spin and orbital for $d^4$ Mott insulators~\cite{PhysRevLett.116.017203}. 
We will give a general and broad
discussion in a forthcoming paper~\cite{Chenunpub}.
	
\end{appendix}

\bibliography{Ref.bib}

\end{document}